\documentclass[useAMS,usenatbib]{mnras}
\usepackage{natbib}
\usepackage{epsfig}
\usepackage{scrtime}
\usepackage{graphicx}   
\usepackage{amsmath}    
\usepackage{amssymb}    
\usepackage{float}
\usepackage[caption = false]{subfig}
\usepackage{multirow}
\usepackage{rotating}
\usepackage{ulem}
\usepackage{booktabs}

\usepackage{soul}

\newcommand{\mincir}{\raise-3.truept\hbox{\rlap{\hbox{$\sim$}}\raise4.truept\hbox{$<$}\ }}

\newcommand{\appropto}{\mathrel{\vcenter{
  \offinterlineskip\halign{\hfil$##$\cr
    \propto\cr\noalign{\kern2pt}\sim\cr\noalign{\kern-2pt}}}}}
\usepackage[dvipsnames]{xcolor}
\usepackage{hyperref}
\usepackage{tikz}

\definecolor{lime}{HTML}{A6CE39}
\DeclareRobustCommand{\orcidicon}{
	\begin{tikzpicture}
	\draw[lime, fill=lime] (0,0) 
	circle [radius=0.13] 
	node[white] {{\fontfamily{qag}\selectfont \tiny ID}};
	\draw[white, fill=white] (-0.0625,0.095) 
	circle [radius=0.007];
	\end{tikzpicture}
	\hspace{-2mm}
}

\foreach \x in {A, ..., Z}{\expandafter\xdef\csname orcid\x\endcsname{\noexpand\href{https://orcid.org/\csname orcidauthor\x\endcsname}
			{\noexpand\orcidicon}}
}

\usepackage[T1]{fontenc}
\usepackage{ae,aecompl}

\usepackage{newtxtext,newtxmath}

\title[Emerging PNe within spirals]{Emerging planetary nebulae within 3D spiral patterns}

\author[V.~Lora~ et al.]{V.~Lora\thanks{E-mail: v.lora@nucleares.unam.mx}$^{1,2}$, J.~A.~Toal\'{a}$^{2}$, J.~I. Gonz\'{a}lez-Carbajal$^{2}$, M.~A.~Guerrero$^{3}$, A.~Esquivel$^{1}$ and G.~Ramos-Larios$^{4}$\\
$^{1}$Instituto de Ciencias Nucleares, UNAM, Apartado postal 73-543, 04510 Ciudad de M\'{e}xico, Mexico\\
$^{2}$Instituto de Radioastronom\'{i}a y Astrof\'{i}sica, UNAM, Antigua Carretera a P\'{a}tzcuaro 8701, Ex-Hda. San Jos\'{e} de la Huerta, 58089 Morelia, Mich., Mexico\\
$^{3}$Instituto de Astrof\'{i}sica de Andaluc\'{i}a, IAA-CSIC, Glorieta de la Astronom\'{i}a s/n, E-18008 Granada, Spain\\
$^{4}$Instituto de Astronom\'{i}a y Meteorolog\'{i}a, CUCEI, Universidad de Guadalajara, Av. Vallarta 2602, 44130 Guadalajara, Jalisco, Mexico
}

\begin{document}
\maketitle
\label{firstpage}

\begin{abstract}
We present the first 3D radiation-hydrodynamic simulations of the formation of planetary nebulae (PNe) emerging from 3D spiral patterns. 
We use the {\sc guacho} code to create 3D spiral structures as a consequence of the distortions on the geometry of the intrinsically isotropic wind of an asymptotic giant branch (AGB) star produced by a companion star in a circular orbit. We found that the orbital period of the binary producing the 3D spiral pattern has consequences on the formation and shaping of the PN itself. Stellar systems with longer period create less entwined 3D spirals, producing PNe with rounder inner cavities, and prevent the expansion of jet towards the polar directions. The spiral fitting procedure used in the literature to predict the binary's orbital period may be misleading in the case of proto-PNe and PNe as spiral patterns are diluted by their own thermal expansion down to the average AGB density profile within a few hundred years and are further disrupted by the action of jets. By adopting a phase of jet ejections between the AGB and post-AGB stages, we are able to recover the morphologies of proto-PNe and PNe that exhibit ring-like structures in their halos.
\end{abstract}


\section{Introduction}
\label{sec:intro}

It is accepted that towards their final stages of evolution, low- and intermediate-mass stars ($M_\mathrm{ZAMS}$=0.8--8~M$_{\odot}$) will create planetary nebulae (PNe). These stars will expel most of their initial mass via dense and slow winds when evolving through the asymptotic giant branch (AGB) phase \citep[$\dot{M}_\mathrm{AGB} \lesssim$10$^{-5}$~M$_\odot$~yr$^{-1}$, $v_\mathrm{AGB} \approx$20~km~s$^{-1}$;][]{Ramstedt2020}. As a consequence of such high mass loss process, their inner hot cores end up exposed. These post-AGB objects are hot enough to produce fast line-driven winds \citep[e.g.,][]{Guerrero2013} and considerably high ionizing UV fluxes that compress and photoionize the previously ejected AGB material to ultimately create what is know as a PN \citep[][]{Kwok2000}.

The variety of morphologies exhibited by PNe \citep[round, elliptical, bipolar, multipolar and irregular; e.g.,][and references therein]{Sahai2011} has been taken as evidence for the action of additional physical shaping mechanisms. These have been proposed to be due to the presence of magnetic fields and/or stellar binarity \citep[or multiplicity; see][]{Zijlstra2015,Akashi2017} and have been extensively tested with high-quality numerical simulations \citep[see, e.g.,][and references therein]{Balick2017,GS2020,Zou2020}.

Within the list of morphological features of PNe, there is one that has gain much attention in recent years: ring-like structures in their halos. These were first reported in high-quality {\it Hubble Space Telescope} images \citep[{\it HST};][]{Terzian2000,Balick2001,RL2012}, however, ring-like structures have also been detected with ground-based  \citep[see][]{Corradi2004} and IR observations \citep[e.g.,][]{Phillips2009,RL2011}. \citet{RL2016} studied a sample of $\sim$650 proto-PNe and PNe and found that the occurrence of ring-like features in the halos of PNe is only about 8 per cent of their analysed sample. These authors classified the ring-like structures as rings, broken rings (arcs), equatorial arcs, disconnected arcs and elliptical arcs.

In addition to the ring-like structures around proto-PNe and PNe, the circumstellar envelopes (CSEs) of some AGB stars exhibit clear spiral patterns. 
The first one was discovered around the carbon star AFGL\,3068 in {\it HST} images \citep[][]{Mauron2006}. Sub-millimetre observations have proven to be even more illustrative due to the possibility to create 3D velocity maps \citep[see, for example][and references therein]{Maercker2012,Doan2020,Homan2018,Randall2020}.

Currently, the most accepted explanation for the formation of spirals around AGB stars is that of binary interactions \citep[e.g.,][]{Harpaz1997}.
Early simulations found that the presence of a companion is able to modify the geometry of the AGB isotropic wind \citep{Mastrodemos1999,He2007}.
In this scenario, spiral patterns appear when the viewing angle is perpendicular to the binary's orbit, whilst arcs and rings are the results of projection effects \citep[see][and references therein]{Kim2019}. Nevertheless, more intricate phenomenology has been presented in the literature \citep[see, e.g.,][]{Aydi2022}. For example, \citet{Malfait2021} showed that the propagation of shocks through the 3D spiral structure might disrupt it, while \citet{Castellanos2021} demonstrated that the wind-interaction of two red giant stars might cause hourglass morphologies. Interestingly, \citet{Decin2020} suggested that spirals around AGB stars can also be shaped by orbiting giant planets as recently tested by \citet{Maes2021}.

It is then reasonable to assume that the ring-like structures in the haloes of PNe are the remains of spiral structures around their precursor AGB stars. 
This idea has been recently tested by \citet{Guerrero2020} using multi-epoch {\it HST} observations of AFGL\,3068 and the PNe NGC\,6543, NGC\,7009 and NGC\,7027 to demonstrate that the ring-like structures in the PNe expand with radial velocities ($\sim$15~km~s$^{-1}$) similar to that of spiral structures around AGB stars. 
\citet{Guerrero2020} further propose that the original AGB spiral pattern is subsequently disrupted by the different effects giving rise to the PNe. In particular the expansion of the fast wind slamming into the slow AGB material produces hydrodynamical instabilities that prevent the material to be uniformly ionized, creating streaks of alternate ionized/neutral material \citep[see][]{Williams1999}. Those illumination effects, in addition to the presence of jet-like ejections, create patterns of incomplete arcs \citep[e.g.,][]{Balick2012,Balick2013,Sahai1998}.

In this paper we present the first numerical study of the early formation of PNe within 3D spiral patterns. We study cases with precessing jets in combination with a fast wind from a post-AGB phase. This paper is organised as follows. In Section~\ref{sec:code} we described the numerical scheme and initial conditions of our simulations. In Section~\ref{sec:results} we present our numerical results, which are subsequently discussed in Section~\ref{sec:discussion}. Our conclusions are presented in Section~\ref{sec:summary}.

\section{Simulations}
\label{sec:code}

\begin{figure*}
    \centering
    {\includegraphics[width=0.9\textwidth]{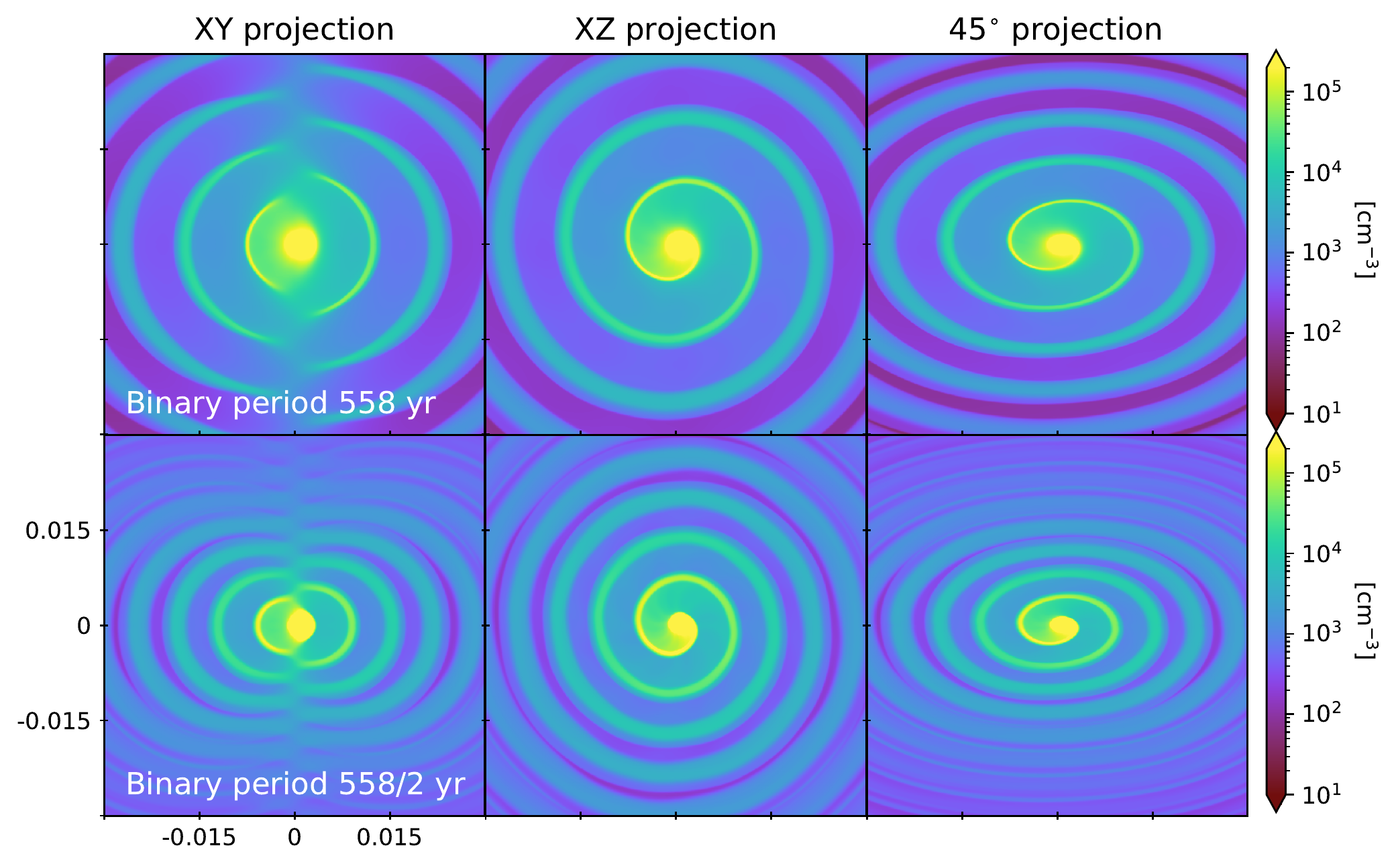} }
    \caption{Number density $n$ distribution of the two initial set ups used in our simulations. The top rows show the results for a binary with a period of 558~yr (Model H1) whilst the bottom rows show results for a period of 558/2~yr (Model H2). Left, middle and right columns show different 2D cuts corresponding to the $x-y$ and $x-z$ planes and a 45$^{\circ}$ inclination angle with respect with the $x-z$ plane. In all cases the orbital plane is coincident with the $x-z$ plane.}
    \label{fig:spirals}
\end{figure*}

We ran our simulations using the radiation-hydrodynamic Eulerian code {\sc guacho} \citep[see][and references therein]{Esquivel2009,Esquivel2013}. 
The code solves the hydrodynamic equations in a uniform Cartesian grid with a second-order accurate Godunov-type method, using a linear slope-limited reconstruction and the HLLC approximate Riemann solver \citep{Toro1994}. 
The {\sc guacho} code includes a modified version of the ionizing radiation transfer presented by \citet{Raga2009} to include the photoionization induced by the central source using a Monte-Carlo ray transfer method  \citep[see][]{Esquivel2013,Schneiter2016}.

The code, which is publicly available\footnote{\url{https://github.com/esquivas/guacho}},
has been appropriately tested and used to model a variety of astrophysical phenomena \citep[e.g.,][]{Estrella2019,Esquivel2019,Lora2021,Schneiter2016}. In particular, our group has recently demonstrated its capabilities in modelling PNe for the case of the born-again PN HuBi\,1 \citep{Toala2021}.

The code solves the rate equation for neutral and ionized hydrogen simultaneously with the Euler equations
\begin{equation}
    \frac{\partial n_\mathrm{HI}}{\partial t} + \nabla \cdot(n_\mathrm{HI} {\bf u})= 
    n_\mathrm{e} n_\mathrm{HII} \alpha(T) - n_\mathrm{HI} n_\mathrm{HII} c(T) - n_\mathrm{HI} \phi,
\end{equation}
\noindent where 
{\bf u} is the flow velocity, 
$n_\mathrm{e}$, $n_\mathrm{HI}$ and $n_\mathrm{HII}$ are the electron, neutral hydrogen and ionized hydrogen number densities, 
$\alpha(T)$ is the recombination coefficient,
$c(T)$ is the  the collisional ionization coefficient and 
$\phi$ is the H photoionization rate due to the central source.

The ionization fraction of hydrogen
\begin{equation}
    \chi=\frac{n_\mathrm{HII}}{n_\mathrm{HI}+n_\mathrm{HII}},
\end{equation}
\noindent where the total number density is $n=n_\mathrm{HI}+n_\mathrm{HII}$, is used to include energy losses from radiation. The implementation used considers the hydrogen ionization fraction and temperature to estimate a non-equilibrium cooling (i.e. cooling by ions out of ionization equilibrium) for temperatures in the range of $10^4<T\lesssim 5\times 10^4~\mathrm{K}$  and it switches to a cooling that depends only on temperature for $T \gtrsim 5\times 10^4~\mathrm{K}$ following the prescription in \citet{Biro1995}. The cooling due to molecular emission, which can be important at low densities and temperatures, is not included. It must be noted, however, that the shocks produced in our models can easily raise the temperature above the range in which molecular cooling is relevant. We also note that neither thermal conduction nor an explicit treatment of dust are included in the simulations presented here.


\subsection{Numerical setup}

All results presented here correspond to simulations ran on a 3D Cartesian grid with a resolution of $(x, y, z)$=(900, 900, 900) on a box of $(0.06\times0.06\times0.06)$~pc$^{3}$ in physical size, that is, a cell resolution $\approx6.67\times10^{-5}$ pc ($\approx$13.7~au). All simulations start with the same initial conditions for the ISM, an initial number density of 1~cm$^{-3}$ with temperature of 100~K.

The simulations are intended to study the formation of PNe within 3D spiral patterns, thus {\sc guacho} is first used to create these structures around a binary system that hosts a mass-losing AGB star. As shown by previous numerical studies (see Sec.~\ref{sec:intro}), the 3D spiral pattern is caused by the orbital motion and the gravitational pull from the stellar companion. The mass of the AGB star and its companion are adopted to be $M_{1}$=0.7~M$_{\odot}$ and $M_{2}$=2.5~M$_{\odot}$, respectively. We adopted the simplest case of a circular orbit with an orbital separation $a$ of 100~au, which translates into a binary period of 558~yr. The orbital plane of the binary system is contained within the $x-z$ plane of the simulation. 
The mass-loss rate of the primary star is set to a conservative value of $\dot{M}_\mathrm{AGB}=5\times10^{-6}$~M$_{\odot}$~yr$^{-1}$ \citep[][]{Ramstedt2020} with a wind terminal velocity of $v_\mathrm{AGB}$=15~km~s$^{-1}$.

\begin{table}
\begin{center}
\caption{Details of the different simulations presented here.}
 \begin{tabular}{ccccc}
\hline
Model    &  AGB binary    &  jet velocity     & Onset of             & Post-AGB\\
label    &  period        &                   & post-AGB$^{\dagger}$ & wind velocity\\
         & $P$            & $v_\mathrm{J}$    & $t_\mathrm{pAGB}$    & $v_\mathrm{pAGB}$   \\
         &  (yr)          &  (km~s$^{-1}$)    & (yr)                 & (km~s$^{-1}$)\\
\hline
H1       & 558     &  \dots            & \dots                       & \dots\\
H2       & 558/2   &  \dots            & \dots                       & \dots \\
\hline
M1       & 558     &  \dots            & 0                           & 1500 \\
M2       & 558/2   &  \dots            & 0                           & 1500 \\
\hline
P2       &  558/2  &  300              & 155                         & 1500\\
P3       &  558/2  &  300              & 210                         & 1500 \\
P4       &  558    &  100              & 155                         & 1500 \\
P5       &  558/2  &  100              & 155                         & 1500 \\
P6       &  558    &  300              & 100                         & 1500 \\
\hline 
\end{tabular}\\
$^{\dagger}$Time period of the interaction of the jet with the 3D spiral just before the onset of the fast, spherically-symmetric post-AGB wind.
\label{tab:models}
\end{center}
\end{table}

The stellar wind is assumed to be isotropic, with a density profile scaled to match the mass loss rate with the terminal velocity $\rho(r) = \dot{M}/(4\pi r^2 v_\infty)$. The wind injection region corresponds to the innermost 10 cells in radius of the numerical grid. Since this region is larger than the wind acceleration region, we do not resolve the launching mechanism, but only the propagation of the wind already at its terminal velocity $v_\infty$. The wind injection region is also larger than the orbit of the binary system, thus the orbit is neither resolved but rather the orbital velocity is added to the wind expansion velocity. The wind injection cells have a temperature of $100~\mathrm{K}$.

We ran a first simulation creating a 3D spiral pattern with the parameters described above, which we labelled as Model H1 (see Table~\ref{tab:models}). 
The evolution of the number density $n$ of Model H1 as seen from the $x-y$ and $x-z$ projections and from a direction 45$^{\circ}$ apart from the $y$ axis 
is presented in the top row of Fig.~\ref{fig:spirals}. For comparison and discussion, we also ran simulations to create a 3D spiral pattern with the same wind properties as those of Model H1 but with a binary period 
shortened by a factor of two, i.e.  558/2~yr. 
This is labelled Model H2. The number density $n$ of the resultant spiral pattern are also shown in Fig.~\ref{fig:spirals} (bottom row).

The spiral patterns presented in Fig.~\ref{fig:spirals} resemble those produced by previous numerical works (see Section~\ref{sec:intro}). In particular, we would like to note the double-shock pattern at the inner and outer edge of the spiral wake produced by the gravitational interaction between the stars in the binary system \citep[see, e.g.,][]{Maes2021}.

\subsection{post-AGB phase}
\label{sec:post_AGB}

\begin{figure}
\centering
\includegraphics[width=\linewidth]{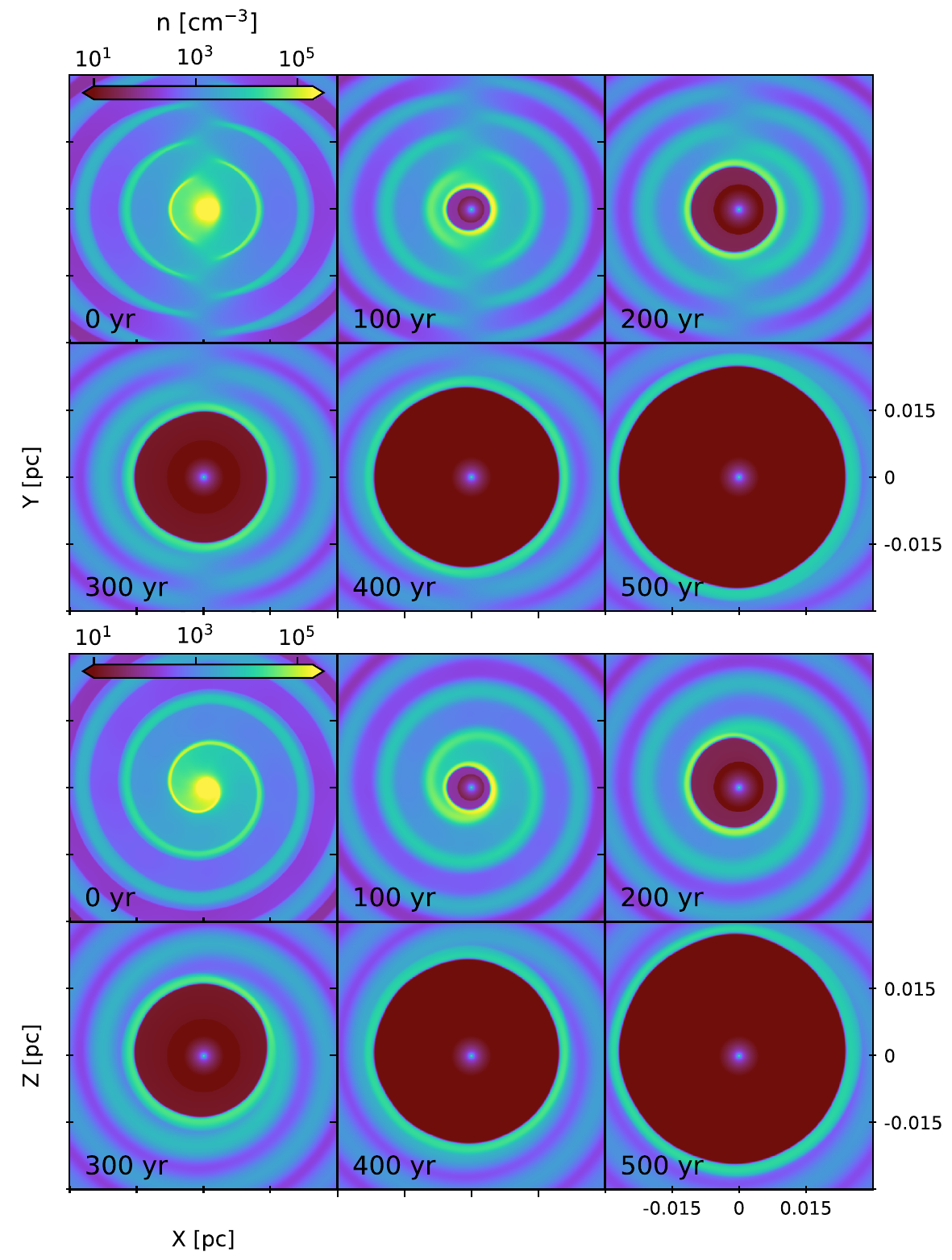} 
\caption{Number density $n$ of the formation and evolution of a PN within a 3D spiral pattern of Model M1 (with binary period 558~yr) produced by a spherically-symmetric post-AGB wind (see Table~\ref{tab:models} for details). 
The top panels show the results for the $x-y$ plane whilst the bottom panels show those for the $x-z$ plane. The results are shown at times $t=0,100,200,300,400$ and $500$ yr after the onset of the post-AGB phase.}
\label{fig:pagb_period}
\end{figure}

\begin{figure}
\centering
\includegraphics[width=\linewidth]{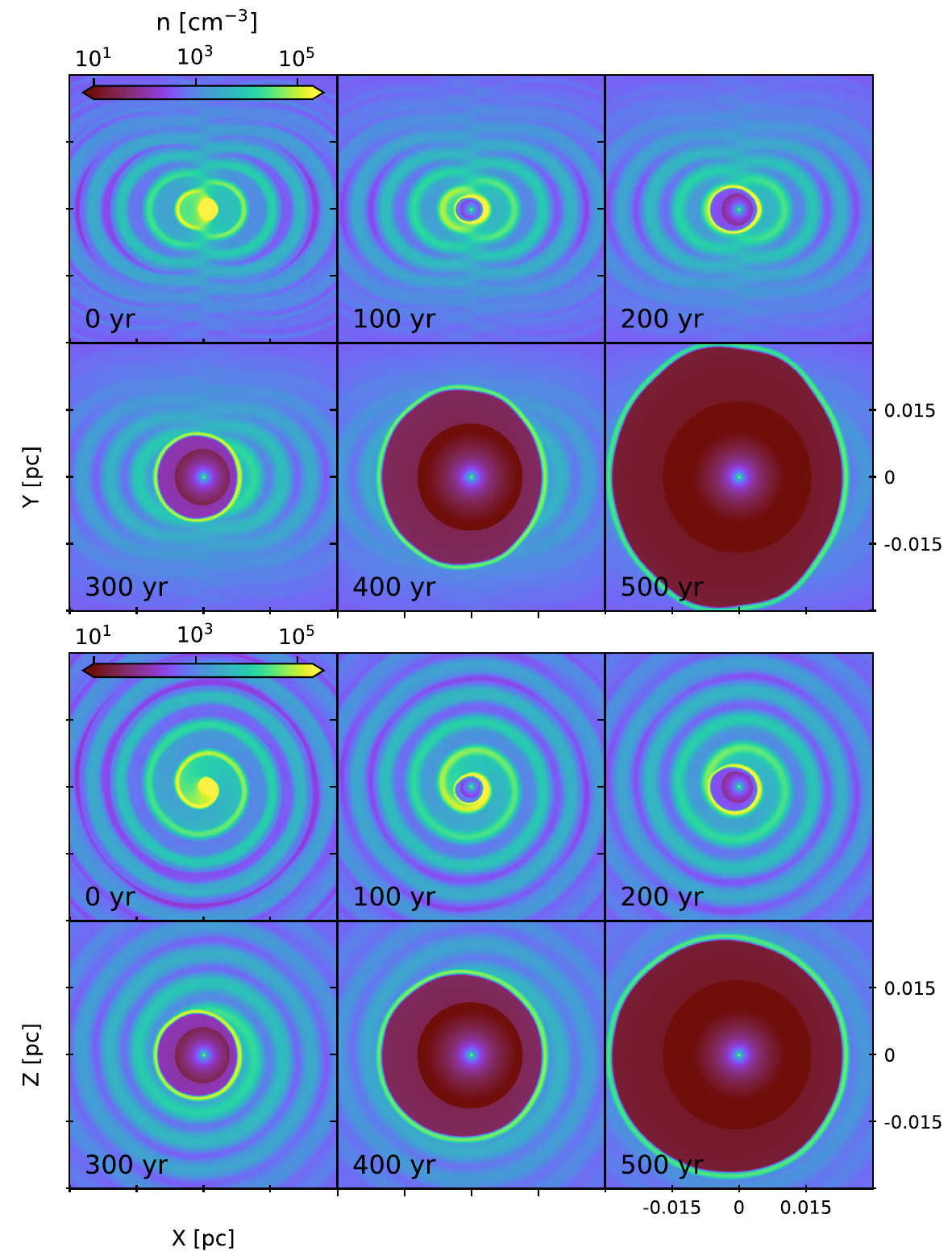} 
\caption{Number density of the formation and evolution of a PN within a 3D spiral pattern of Model M2 (with binary period 558/2~yr) produced by a spherically-symmetric post-AGB wind (see Table~\ref{tab:models} for details). 
The top panels show the results for the $x-y$ plane whilst the bottom panels show those for the $x-z$ plane. The results are shown at times $t=0,100,200,300,400$ and $500$ yr after the onset of the post-AGB phase.}
\label{fig:pagb_period2}
\end{figure}

After the AGB 3D spiral fills the whole computational range, we follow different post-AGB evolution sequences. The simplest case is that of a spherically-symmetric post-AGB stellar wind. 
The stellar wind parameters in the post-AGB phase are set to $\dot{M}_\mathrm{pAGB}=10^{-7}$~M$_\odot$~yr$^{-1}$ and $v_\mathrm{pAGB}=1500$~km~s$^{-1}$ with an ionizing photon flux\footnote{The ionizing photon flux used here is consistent with stellar evolutionary models of solar-like stars with a range of initial masses \citep[see, e.g.,][]{Villaver2002a,Toala2014}.} of $\dot{S}=5\times10^{46}$~s$^{-1}$. The variation of the stellar wind parameters in combination with the high ionizing flux will produce a PN within the 3D spirals. The formation and evolution of a PN by the interaction of a spherically-symmetric post-AGB wind with the 3D spirals produced by models H1 and H2 are presented as Models M1 and M2. The evolution with time of the number density $n$ of Models M1 and M2 are illustrated in Figs.~\ref{fig:pagb_period} and \ref{fig:pagb_period2}, respectively.

We note that a more realistic simulation would include the transition of the stellar wind parameters of the AGB into the post-AGB according to stellar evolution models. However, previous numerical studies have demonstrated that this transition is so short that it has a negligible impact on the evolution of the PN \citep[e.g.,][]{Perinotto2004}. Simulations following the time-dependent evolution of the stellar wind parameters will be pursuit in a subsequent paper.

Given the occurrence of multipolar morphological features in proto-PNe and PNe, in particular in those exhibiting ring-like features \citep[see][and references therein]{Guerrero2020,RL2016}, we also included jets into our simulations. Jets are injected as an intermediate phase between the AGB and post-AGB phases. That is, after the 3D spiral pattern is formed, the jet is injected with a certain duration and finally, the spherically-symmetric post-AGB wind is injected into the numerical domain.

\begin{figure}
\centering
\includegraphics[width=0.8\linewidth]{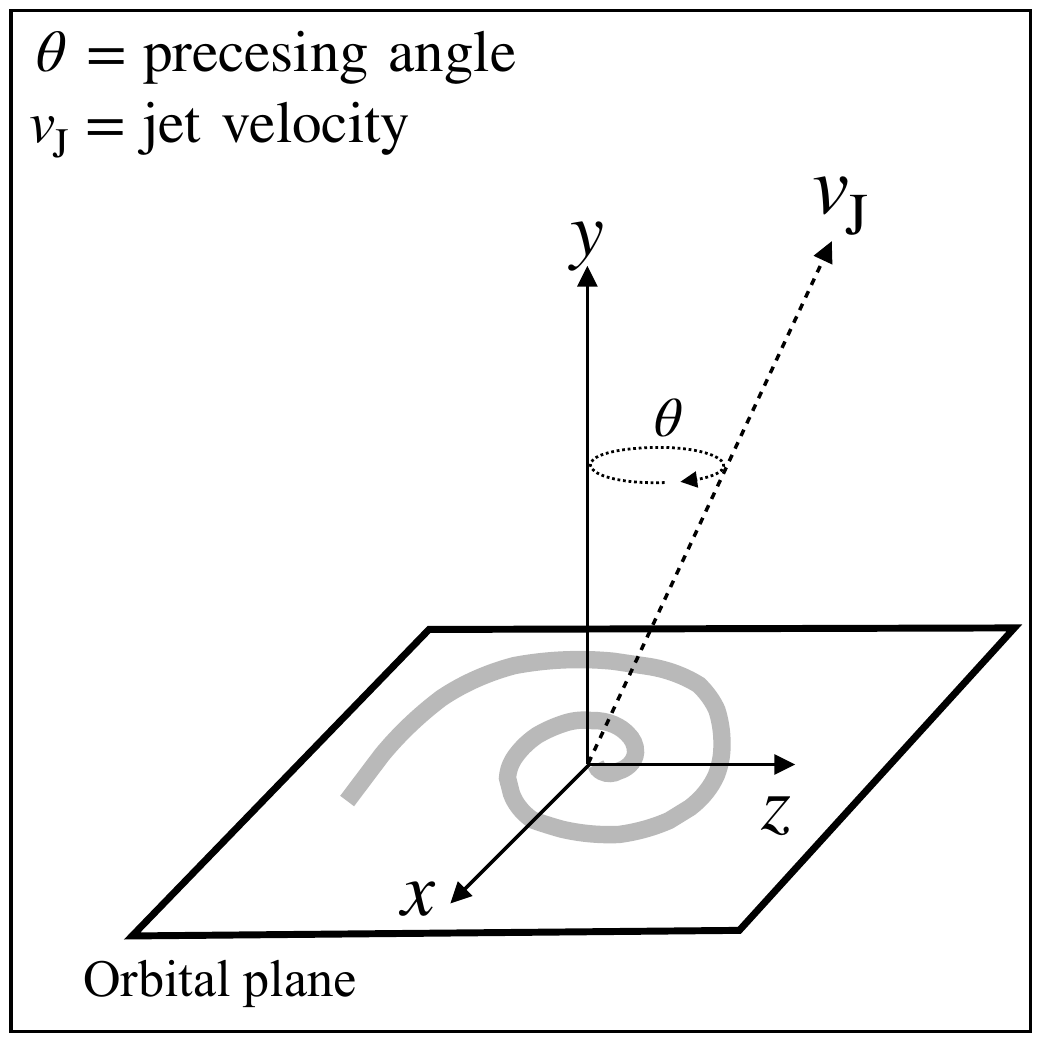}
    \caption{Schematic view of the jet in the simulations. The jet is launched with velocities of $v_\mathrm{J}$=100 and 300~km~s$^{-1}$ with a precessing angle $\theta$=35$^{\circ}$. The orbital plane is coincident with the $x-z$ plane.}
    \label{fig:jet}
\end{figure}

We imposed the jet to have a top-hat cross-section and an opening angle equal to zero degrees (i.e., a pencil-like jet). The jet injection cell zone is defined as a cylindrical region with radius $150~\mathrm{au}$ and height $300~\mathrm{au}$. The jet precesses with an opening angle of 35$^{\circ}$ with respect to the $y$ axis. We note that precession can be produced by the interaction of the binary system with the accretion disk that originates the jet \citep[see][]{Terquem:99}. Its mass-loss rate was set to $\dot{M}_\mathrm{J}=2.5\times10^{-6}$~M$_\odot$~yr$^{-1}$ which is typical for jets in post-AGB systems \citep[e.g.,][]{Bollen2020}. However, we used two different values for the jet velocity: $v_\mathrm{J}$=100~km~s$^{-1}$ which represents about 70 per cent of the PNe harbouring jet-like features \citep{Guerrero2020b}, and 300~km~s$^{-1}$ for the less numerous cases. The precessing period of the jet was set to 320~yr with an ejection period to 80~yr. We note that these values are arbitrary and serve for illustration purposes. The ionizing photon flux is not included during the jet injection. A schematic view of the jet is presented in Fig.~\ref{fig:jet}.

\begin{figure*}
\centering
\includegraphics[width=0.8\linewidth]{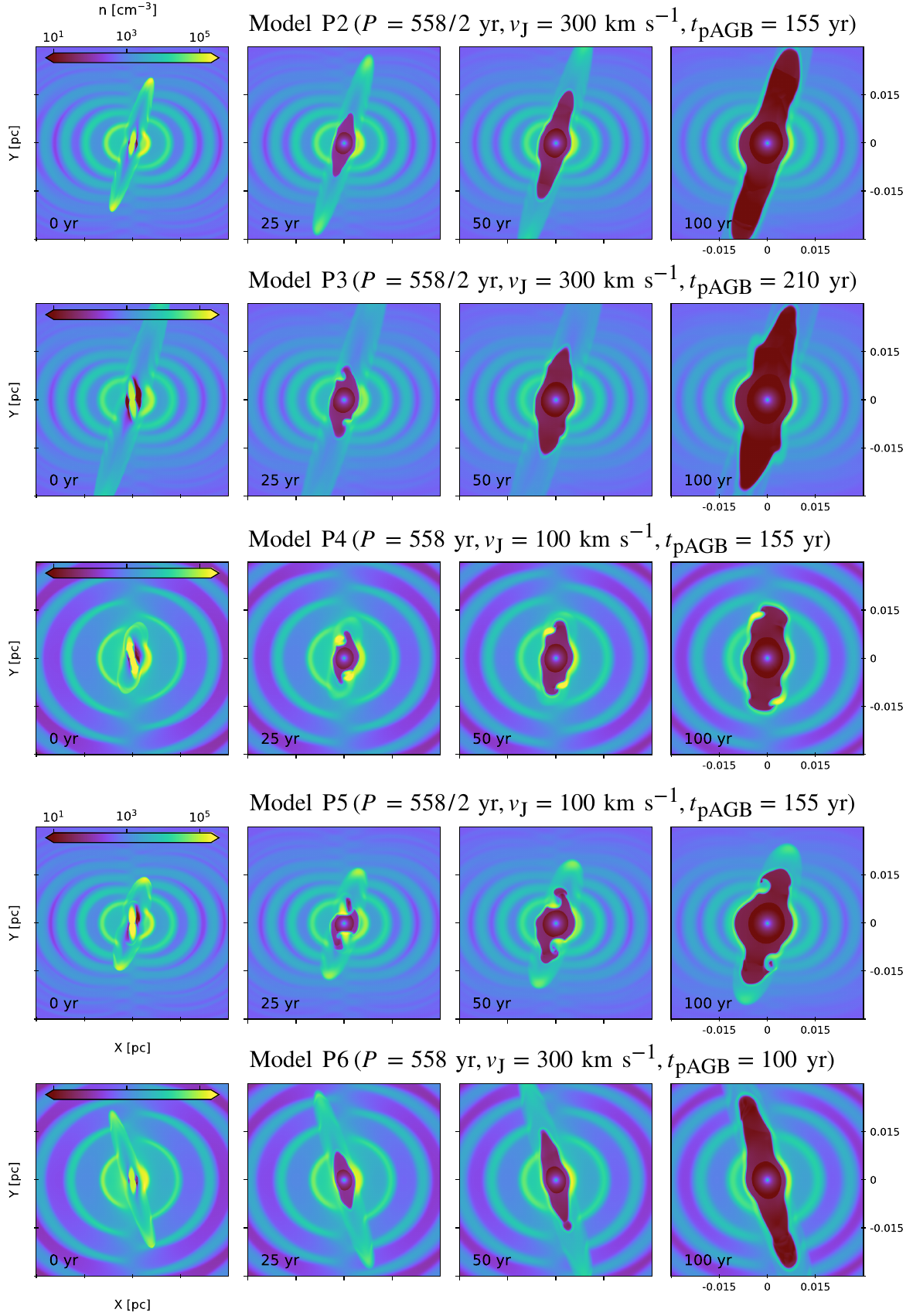} 
\caption{Evolution with time of the number density $n$ for Models P2--P6 as seen in the $x-y$ plane. The results show the integration times $t=0,25,50$ and $100$ yr after the onset of the post-AGB phase. In all cases a jet has been injected and interacted with the 3D spiral pattern before the onset of the post-AGB phase. Details are listed in Table~\ref{tab:models} and described in Sec.~\ref{sec:post_AGB}.}
\label{fig:P2}
\end{figure*}

We ran simulations with different combinations of the AGB spiral period, jet velocity and the time of the onset of the post-AGB phase. The details of the different models including jets labelled from P2 to P6 are listed in Table~\ref{tab:models} and presented in Figure~\ref{fig:P2}. For example, the top and second rows of Fig.~\ref{fig:P2} present the numerical results of a jet with a velocity of $v_\mathrm{J}$=300~km~s$^{-1}$ that expanded into a 3D spiral originated in a binary period of 558/2~yr with a post-AGB phase initiated 155~yr (model P2) and 210~yr (model P3) after the jet onset, respectively. Results from models P4--P6 are presented in other rows in Fig.~\ref{fig:P2}.

\section{Results}
\label{sec:results}

\subsection{Spherical cases}
\label{sec:spherical_cases}

Models M1 and M2 illustrated in Figs.~\ref{fig:pagb_period} and \ref{fig:pagb_period2}, respectively, show the results of the formation of a PN by the injection of a spherically-symmetric post-AGB wind within 3D spiral structures. These figures show the known formation scenario of PNe where the fast post-AGB wind slams the slow AGB wind producing an adiabatically-shocked hot bubble \citep[e.g.,][]{Villaver2002a,Perinotto2004,Stute2006,Toala2014}. The hot bubble, traced in Figs.~\ref{fig:pagb_period} and \ref{fig:pagb_period2} by the low density region, pushes out the previously formed 3D spiral, producing an expanding inner rim. Maps of the temperature structure of the simulations are presented in Fig.~\ref{fig:pagb_period_temp} and \ref{fig:pagb_period2_temp} of Appendix~\ref{sec:app}.

\begin{figure}
\centering
\includegraphics[width=\linewidth]{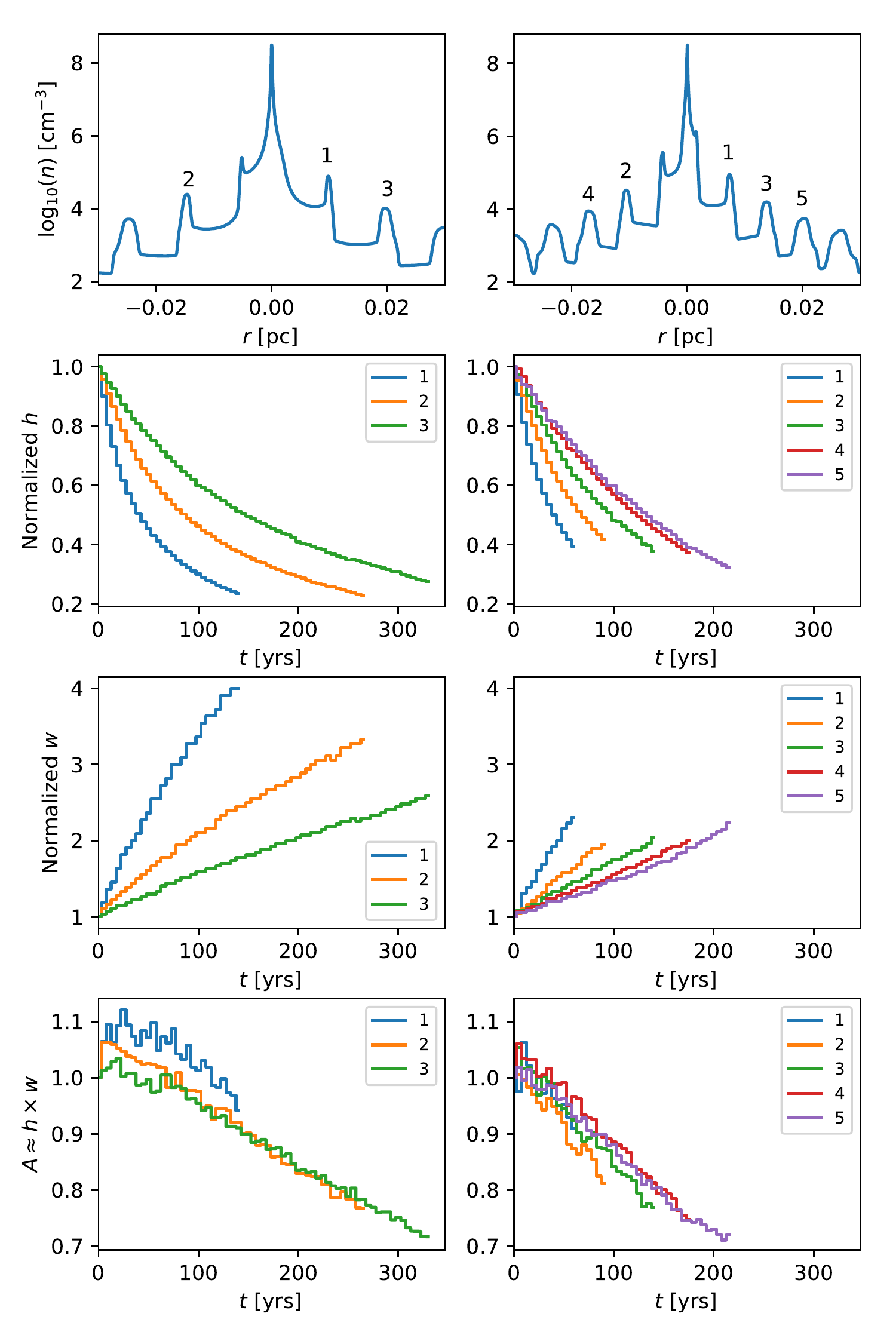}
    \caption{Time evolution of the spiral patterns of simulations M1 and M2 presented in Fig.~\ref{fig:pagb_period} and \ref{fig:pagb_period2}, respectively. The top panels show a density cut in the $x-y$ plane just before the onset of the post-AGB phase. The second, third and last rows show the evolution of the height ($h$), width ($w$) and area ($A\approx h\times w$) of the spiral arm features. Colours represent the selected features labelled on the top panels.}
    \label{fig:rings_evolution}
\end{figure}

\begin{figure}
\centering
\includegraphics[width=0.9\linewidth]{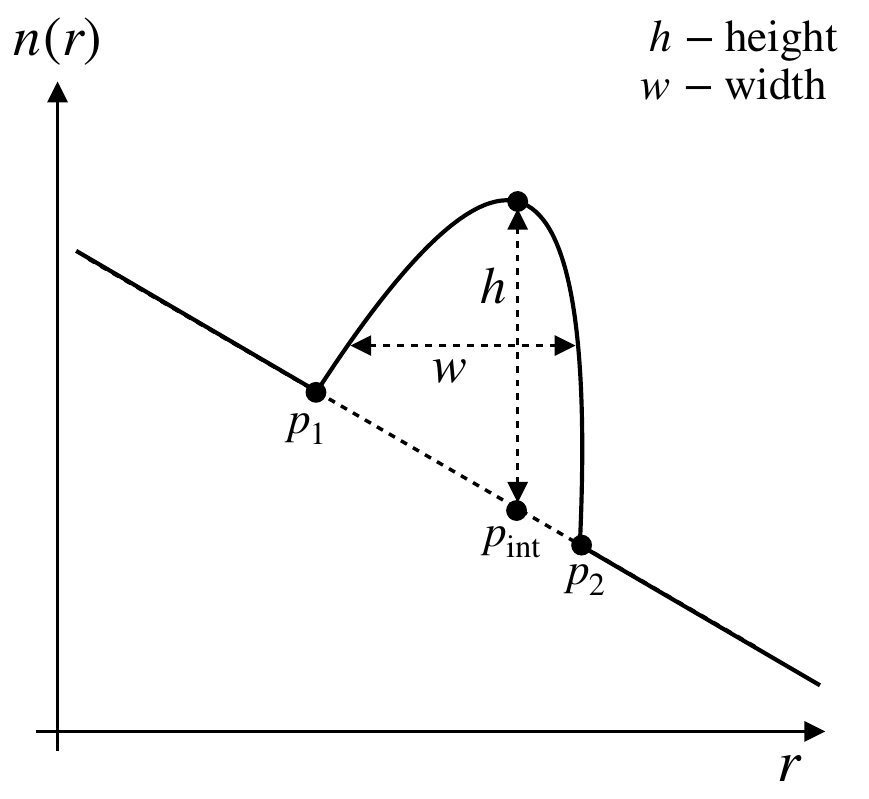}
    \caption{Definition of the height ($h$) and width ($w$) of each spiral feature. $p_1$ and $p_2$ represent the two ends of the peak and $p_\mathrm{int}$ is the linear interpolation point at the peak of the feature. $w$ is defined as the FWHM using $h$.}
    \label{fig:spiral_profile}
\end{figure}

Fig.~\ref{fig:pagb_period} and \ref{fig:pagb_period2} show that the spirals change with time as the PN is formed and evolves. The most obvious effect is the progressive smoothing experienced by the spiral arms, which reflects their dilution effect with time. To quantify this effect, we study the time evolution of the height and width of the spirals features. We created density cuts in the $x-y$ plane at different times, starting at the onset of the post-AGB phase.

In the top panels of Figure~\ref{fig:rings_evolution} we show the initial density profile extracted from the spiral patterns of Models M1 and M2 in the top rows of Fig.~\ref{fig:pagb_period} and Fig.~\ref{fig:pagb_period2}. 
First, we note that the spiral arm features are mounted onto a $n \propto r^{-2}$ density profile more or less typical for the distribution of the AGB wind \citep{Villaver2002b}.

We have selected several spiral maxima structures, labelled from 1 to 5 in the top panels of Fig.~\ref{fig:rings_evolution}, to follow their evolution. For this, we will define their width ($w$) and height ($h$) as illustrated in Figure~\ref{fig:spiral_profile}. We start by searching the interpolation point ($p_\mathrm{int}$) between the two ends of each spiral arm feature, defined by $p_1$ and $p_2$, which corresponds to the peak of the feature. $h$ is then defined as the difference between the peak density of the spiral arm and the density at $p_\mathrm{int}$, and $w$ as the full width at half maximum (FWHM) using $h$. This procedure was performed for each spiral arm structures labelled in the top panel of Fig.~\ref{fig:rings_evolution} through their evolution. The height and width are normalized to the initial values just before the onset of the post-AGB phase.

The time evolution of the parameters $w$ and $h$ of the peaks are shown in the second and third rows of Fig.~\ref{fig:rings_evolution}, respectively, for models M1 (left) and M2 (right). The height $h$ diminishes exponentially with time, depending on the spiral feature one follows. For example, spiral features closer to the origin reduce their $h$ faster than those originally placed at larger distances (see second row of Fig.~\ref{fig:rings_evolution}). This might be attributed to the compression produced by the inner rim of the PNe expanding onto the spiral pattern. Meanwhile, the width $w$ increases nearly linearly with time.

Using these two quantities, we defined an effective area of each spiral pattern as $A= w \times h$. This is shown to reduce inversely proportional to time in the bottom row of Figure~\ref{fig:rings_evolution}. In fact, this result can be also extended to argue that the volume of the spiral patterns is also reducing with time (taking any arbitrary spiral pattern length).

This analysis suggests that the inner structure of the spiral pattern is diluting in relatively small time scales. In particular, we found that any feature closer than 0.02~pc will be diluted within the density profile to 20 per cent of their original contrast within $\lesssim$300~yr of evolution. This result might be a combination of effects. 
To start with, the formation of the hot bubble makes it expand into the 3D spiral forming an inner rim of swept material. Secondly, as the result of the spiral own thermal pressure. 
We note that \citet{Maes2021} demonstrated that the separation of the inner and outer edges of the spiral pattern is caused by the velocity dispersion of the gas within this structure.
We also note that our spiral patterns are more simplistic than those obtained by \citet{Maes2021} by varying different parameters in their simulations (such as wind velocity and orbital properties). Further exploration of the destruction of spiral patterns would require a similar approach as those authors, which is beyond the scope of the present paper.

Another effect worth mentioning is that of Model M2, which exhibits the formation of an elliptical inner cavity in the $x-y$ plane (Fig.~\ref{fig:pagb_period2}). It is natural to expect that an elliptical PN will be subsequently formed when injecting a spherically-symmetric post-AGB wind into the 3D spiral structure, because most of the material is located in the orbital plane. The pressure of the hot bubble will help pushing towards the polar directions of the 3D spiral structures where the CSE pressure is smaller \citep[see, e.g.,][]{Malfait2021}. These results suggest that the period of the binary producing the spiral pattern in the CSE can also affect the PN shaping. The less entwined spirals produced by longer binary orbital periods (e.g., Model M1; Fig.~\ref{fig:pagb_period}) result in more isotropic CSE than those formed around binary systems with shorter orbital periods (e.g., Model M2; Fig.~\ref{fig:pagb_period2}).

\subsection{Jets$+$post-AGB wind models}

Fig.~\ref{fig:P2} shows the different numerical results of models P2--P6 described in Table~\ref{tab:models}. These correspond to number density maps of models that include jet ejections followed by a fast post-AGB wind. Depending on the specific time of the onset of the post-AGB phase the jets are able to produce smaller (or larger) bipolar cavities expanding into the 3D spiral pattern.

The interaction of the jets with the 3D spiral produces tunnel-like shocked structures with temperatures of $\sim10^{4}$~K, similarly to what is observed in proto-PNe \citep[see the case of CRL\,618 in][]{Balick2013}. Examples of the different configurations can be seen in the leftmost columns ($t=0$~yr) of Fig.~\ref{fig:P2}, which show the $x-y$ plane. At this point, the simulations might be visually compared to proto-PNe, although we note that we are not tailoring the results to any specific object. When the fast, spherically-symmetric post-AGB wind is injected into the computational grid it creates hot bubbles that preferentially fill the tunnel-like structures produced by the jets. This effect is illustrated in Fig.~\ref{fig:P2_temp} in Appendix~\ref{sec:app}.

\begin{figure*}
\centering
\includegraphics[width=\linewidth]{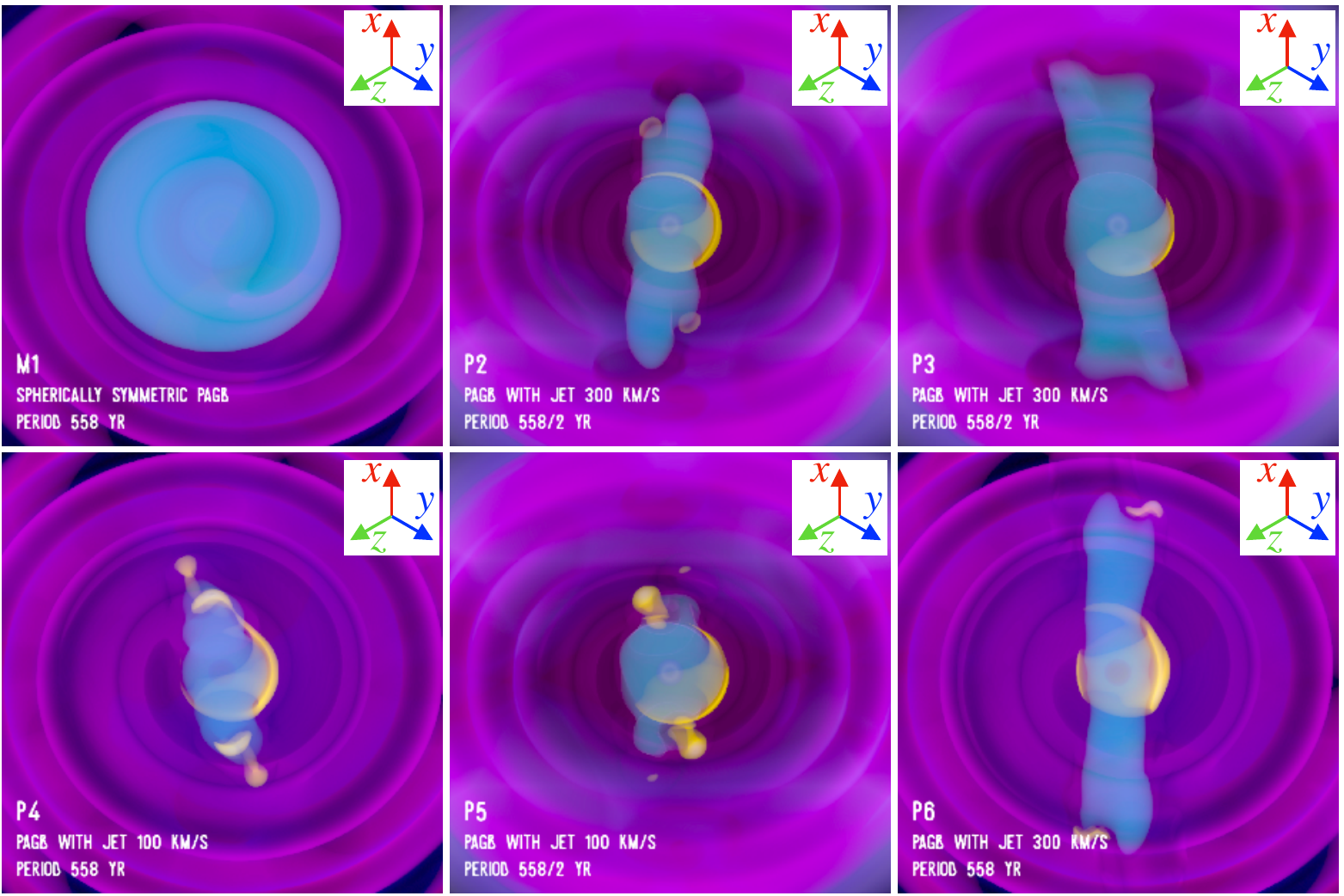}
\caption{Examples of synthetic nebular images obtained for our simulations. All panels correspond to models rotated $\pi$/4 over the three axis. Each panel shows the results from different set ups. Four different tinges are used to illustrate different density ranges: blue [1--10$^{2}$]~cm$^{-3}$, purple [10$^{2}$--10$^{3}$]~cm$^{-3}$, magenta [10$^{3}$--10$^{4}$]~cm$^{-3}$ and yellow [10$^{5}$--10$^{6}$]~cm$^{-3}$.}
\label{fig:renders}
\end{figure*}

In addition, similarly to what we observe in the simulations without jets, the case with the spiral formed with the longer period (558~yr) does not allow the jet to expand to larger distances. This effect can be clearly seen by comparing, for example, models P4 and P5 in the third and fourth rows of Fig.~\ref{fig:P2}, respectively. In both models the jet velocity was set to $v_\mathrm{J}$=100~km~s$^{-1}$ and the post-AGB wind started after 155~yr of evolution of the jet. The only difference is that model P4 was produced by a binary period of 558~yr and the spiral in model P5 with that of 558/2~yr. In model P4 the bipolar structure inflated by the fast wind has reached to distance of 0.015 pc from the central star after 100~yr of evolution, while model P5 has almost left the computational domain (0.03 pc) in the same amount of time. Furthermore, model P5 exhibits a double-lobed young PNe different to that of model P4.

\section{Discussion}
\label{sec:discussion}

If one accepts that most PNe are formed somehow by the evolution of binary systems, ring-like structures are to be expected in their halos. \citet{Corradi2004} raised this issue and estimated that at least 35 per cent of PNe should exhibit these structures. However, the statistical work presented by \citet{RL2016} showed that only 8 per cent of 650 proto-PNe and PNe exhibit ring-like features in their halos. Such low occurrence of ring-like structures might suggest the existence of physical mechanism erasing their signature.

Our numerical results are consistent with the low occurrence of spirals and ring-like features. In accordance with previous findings, we found that the spiral structure can disappear in relatively short periods of time owing to its broadening, initially caused by dynamical interactions \citep[see][]{Maes2021}. In addition, the densest and brightest inner structures will be swept up by the early expansion of the inner rim or the presence of jets. The combination of these effects will reduce the detection of spirals or any ring-like feature in the halos of proto-PNe and PNe.

Other effects that might help destroying the spiral structure is the uneven ionization patterns produced by the shadow instabilities formed as a consequence of the turbulence in the wind-wind interaction zone at the inner rim of evolved PNe. Given the numerical resolution of the present simulations, these are not resolved. However, these effects were explored in the high-resolution 2D radiation-hydrodynamic simulations presented by \citet{Guerrero2020} for longer evolution times than those explored in the present work. Thus, we suggest that trying to assess the orbital parameters from spirals and/or ring-like structures from proto-PNe and PNe, as done for AGB stars \citep[see, e.g.,][]{Kim2012}, should be taken with extreme care. If these structures have been strongly affected, they would not represent accurately the properties of the orbital parameters. For example, the case of NGC\,7027 where multiple disconnected arc-like features in addition to jets are detected surrounding its main cavity \citep[e.g.,][]{Kastner2020,Moraga2023}. 
We note that similar structures have been unveiled by the unprecedented images of NGC\,3132 obtained by the {\it James Webb Space Telescope} \citep{DeMarco2022}\footnote{see \url{https://science.nasa.gov/webbs-southern-ring-nebula}}.

We do not explore a wide range of orbital parameters for the spiral pattern formation as performed in \citet{Kim2019}, but our simulations suggest that more entwined spirals help shaping bipolar hot bubbles and PNe as they produce less density in the polar regions away from the binary's orbital plane. The effect is not notorious when a spherically-symmetric fast post-AGB wind is injected into the spiral, but it largely affects the propagation of bipolar structures carved by jets. This is well illustrated by simulations P4 and P5 (third and fourth rows in Fig.~\ref{fig:P2}), with longer and shorter binary periods, respectively.
Apparently the more efficient rearrangement of the AGB wind towards the orbital plane on systems with shorter binary periods (P4) favors the expansion of outflows along the polar direction.

Recently, \citet{Malfait2021} showed that some departures from a clear 3D spiral pattern can be caused by varying the AGB wind and binary eccentricity. These authors showed that an AGB wind as slow as  5--10~km~s$^{-1}$ and orbital eccentricities of 0.25 and 0.50 produce gas compression close to the companion creating irregular spirals. Furthermore, these authors also found that bipolar outflows form given their inefficient cooling near the companion. We stress that the formation of PNe in such messy structures might help forming irregular PNe. This is an interesting issue that will be worth pursuing in future simulations.

\subsection{Comparison with observations}

A significant fraction of the proto-PNe and PNe with ring-like features reported by \citet{RL2016} exhibit bipolar or multipolar structures evidently departing from spherical symmetry. It is expected that accretion onto the late AGB's companion produces a bipolar ejection \citep[e.g.,][]{Manick2021} disrupting the 3D spiral structure. These effects have been detected in the AGB star $\pi^{1}$~Gruis using sub-millimeter ALMA observations \citep{Doan2020} where a bipolar ejection is being produced in the polar direction from the orbital plane and its 3D spiral structure. Furthermore, precessing jets might be also present in more evolved objects, for example, GLMP\,870, PM\,1-176, and PM\,2-42 \citep{RL2016}.

\begin{figure*}
\centering
\includegraphics[width=\linewidth]{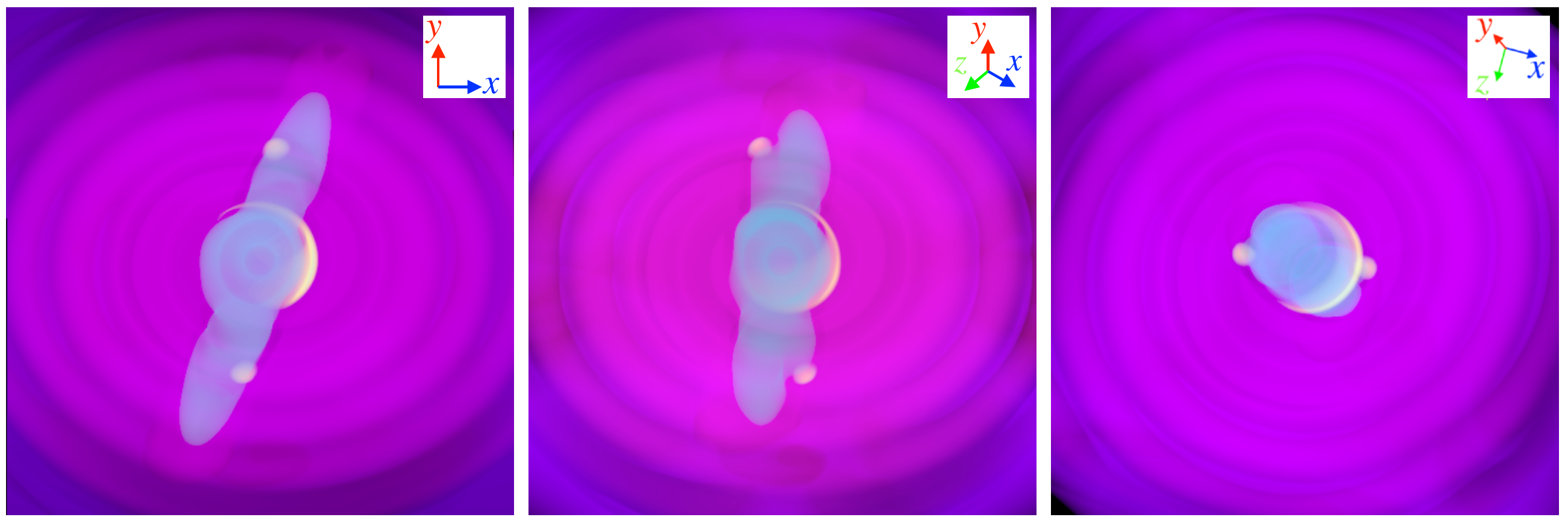}
\caption{Synthetic images of model P2 produced at 80~yr after the post-AGB phase evolution. Each panel shows different angle rotations. Four different tinges illustrate different density ranges: blue [1--10$^{2}$]~cm$^{-3}$, purple [10$^{2}$--10$^{3}$]~cm$^{-3}$, magenta [10$^{3}$--10$^{4}$]~cm$^{-3}$ and yellow [10$^{5}$--10$^{6}$]~cm$^{-3}$.}
\label{fig:renders2}
\end{figure*}

To compare our simulations with observations, we created density render images to mimic nebular properties. This was done by making use of the Multi-code Analysis Toolkit for Astrophysical Simulation Data {\sc yt} \citep{Turk2011}. This software allowed us to create images by casting rays through the 3D volume and integrating the radiation transfer equations. The transfer function is chosen with transparency and colours that depend on the value of the field that is being rendered. Examples of these, with the transparency varying linearly with density (denser is more opaque), are presented in Figures~\ref{fig:renders} and \ref{fig:renders2}. We used four different tinges to illustrate different density ranges from 1 to 10$^{6}$~cm$^{-3}$.

The examples presented in Fig.~\ref{fig:renders} present randomly-selected times to illustrate the different morphologies produced by the models, these were produced by applying a rotation of $\pi/4$ in the three axis. The top left panel in Fig.~\ref{fig:renders} shows the simulation of a spherically-symmetric fast wind expanding into a spiral produced by a 558~yr orbital period (Model M1). The other panels show synthetic images from Models P2--P6.

Without tailoring any of our simulations to specific objects, our synthetic images reproduce some properties of observed PNe. 
In particular, the image in the bottom left panel of Fig.~\ref{fig:renders}, corresponding to model P4, presents density enhancements (in yellow) indicative of the interactions between a precessing jet and the 3D spiral structure reminiscent of the bow-shock-like structures observed in NGC\,6543 \citep{Balick2001,RL2016} and to a lesser extent in NGC\,7009 \citep{Guerrero2020}. The interactions of jets with the 3D spiral structure also produce density enhancements in models P2 and P6 in Fig.~\ref{fig:renders}.

Our synthetic images corroborate previous findings that, depending on the viewing angle, the spiral pattern seems to vanish and produce apparent ellipses or disconnected ring-like structures. Images extracted from models P2, P3 and P5 shown in Fig.~\ref{fig:renders} are examples of this, where projection effects make the 3D spiral look like elliptical concentric structures. Interestingly, these cases are the results of simulations adopting an orbital period of 588/2~yr, suggesting that smaller orbital periods might help hiding the 3D spiral structure more easily even when viewed through the orbital plane. 
Apparently the more entwined spiral arms resulting from binaries with smaller orbital periods imply lower arm to inter-arm density contrasts that hamper the detection of these arms at non-preferential viewing angles.

To further illustrate projection effects, we present in Fig.~\ref{fig:renders2} different synthetic images for model P2 obtained by varying the viewing angle. The apparent PN morphology is obviously affected by the viewing angle, going from highly elongated to quadrupolar and to mostly round. This is also the case for the 3D spiral pattern, going from a spiral to a series of concentric ellipses in the middle panel of Fig.~\ref{fig:renders2}.

Observations of the extended molecular emission around AGB stars have been useful disclosing the true 3D spiral structure of their density distributions using its radial velocity component, but this has not been the case of proto-PNe and PNe, which are mainly detected in optical and IR images without radial velocity information. The expansion velocity has only been measured for some ring-like structures around PNe using expansion patterns from multi-epoch observations \citep[see][and references therein]{Guerrero2020}.
Cases such as IC\,418, IC\,4406, NGC\,3242, NGC\,6543 and NGC\,7009 \citep{Balick2001,RL2011,RL2012,RL2016,Guerrero2020} exhibit more or less concentric density structures that, according to the results presented here, are reminiscent of spiral patterns. We conclude that in the cases where well-defined concentric structures are observed, it is not possible to infer orbital properties such as the orbital plane orientation, unless projection effects are somehow disentangled, yet the separation between ring-like structures still provides a rough estimate of the binary period.

\section{Summary} 
\label{sec:summary}

We presented 3D radiation-hydrodynamic numerical simulations produced with the {\sc guacho} code of the early phases of formation of PNe emerging from 3D spiral patterns. 
In our simulations, the 3D spiral is formed as the result of the effects on the geometry of the originally isotropic AGB wind induced by the orbital motion and gravitational pull of a companion assumed to be in a circular orbit. Subsequently, we simulate the post-AGB phase accounting for i) jet-like ejections and ii) a spherically symmetric fast wind. 
Our results can be summarised as follows:
\begin{itemize}
    
\item 
In our simplest case, the spherically-symmetric fast wind expands compressing the previously-shaped 3D spiral pattern. We estimate that the densest regions of the spiral will disappear in a few $\lesssim$500~yr owing to its natural thermal expansion down the density profile left by the AGB star. The surviving outer structure of the spiral pattern will be further disrupted by the uneven ionization produced by the shadowing instability, which creates streaks of alternate ionized material. Jets will also contribute to the disruption of the 3D spiral pattern. Trying to assess orbital parameters from disrupted spirals, that is ring-like structures, in the halos of PNe should be taken with care, specially for those cases in which the structures have been heavily disrupted such is the case of NGC\,7027.

\item 
We found that the orbital period has consequences in the early formation of PNe. In particular, we show that more entwined 3D spirals produced with smaller orbital periods hinder the expansion of hot bubbles towards the orbital plane, allowing the expansion of elliptical cavities towards polar directions. This affects notably to jets injected close to the direction orthogonal to the orbital plane, progressing throughout lower density material in systems with shorter orbital periods, thus expanding up to further distances.

\item Although we do not tailored our simulations to any specific object, we are able to explain different observed PNe that exhibit ring-like structures. In particular, we found that NGC\,6543 resembles morphologically to the results from model P2. This model exhibits the presence of bow-shock-like features produced by the interaction of the jets with the densest regions of the 3D spiral. This effect has been discussed previously to be causing the morphology of the jet-like features in NGC\,6543 observed with the {\it HST}.

\end{itemize}

\section*{Acknowledgements}

The authors are grateful to the referees that helped improve the presentation and discussion of our results. VL gratefully acknowledges support from the \mbox{CONACyT} Research Fellowship program. JAT thanks Fundación Marcos Moshinsky (Mexico) and the Direcci\'{o}nn General de Asuntos del Personal Acad\'{e}mico (DGAPA) of the Universidad Nacional Aut\'{o}noma de M\'{e}xico (UNAM) project IA101622. JIG-C thanks Consejo Nacional de Ciencia y Tecnolog\'{i}a (Mexico) for a scholarship grant. MAG acknowledges support of grant PGC 2018-102184-B-I00 of the Ministerio de Educación, Innovación y Universidades cofunded with FEDER funds. GR-L acknowledges support from Universidad de Guadalajara, CONACyT grant 263373 and Programa para el Desarrollo Profesional Docente (PRODEP, Mexico). This work has made extensive use of NASA's Astrophysics Data System.

\section*{Data availability}
The results of the simulations presented in this work will be shared on reasonable request to the first author.

\appendix

\section{Temperature maps}
\label{sec:app}

\begin{figure}
\centering
\includegraphics[width=\linewidth]{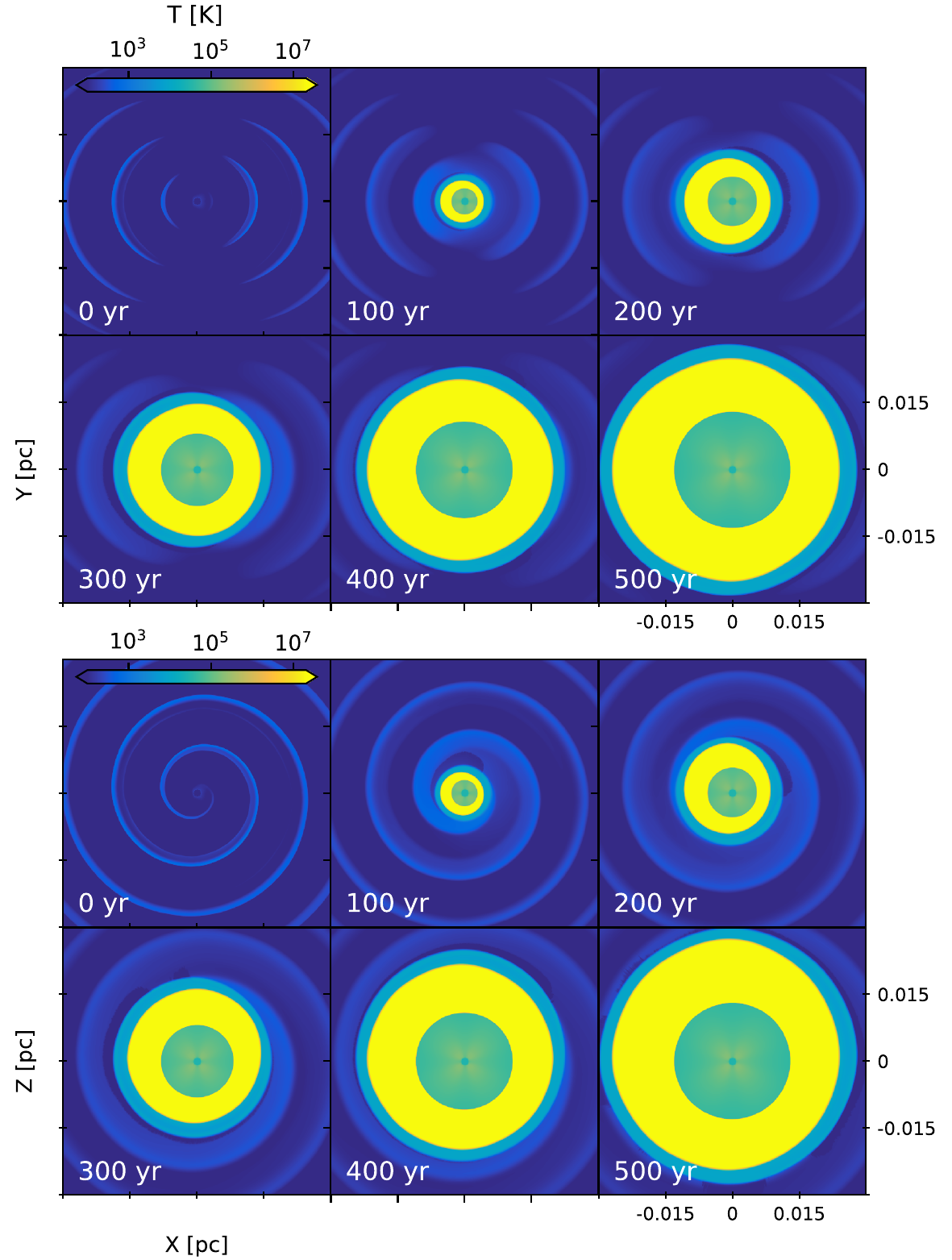}
    \caption{Temperature $T$ of the formation and evolution of a PN within a 3D spiral pattern of Model M1 (with binary period 558~yr) produced by a spherically-symmetric post-AGB wind (see Table~\ref{tab:models} for details). The top panels show the results for the $x-y$ plane whilst the bottom panels show those for the $x-z$ plane. The results are shown at times $t=0,100,200,300,400$ and $500$ yr after the onset of the post-AGB phase.}
\label{fig:pagb_period_temp}
\end{figure}

\begin{figure}
\centering
\includegraphics[width=\linewidth]{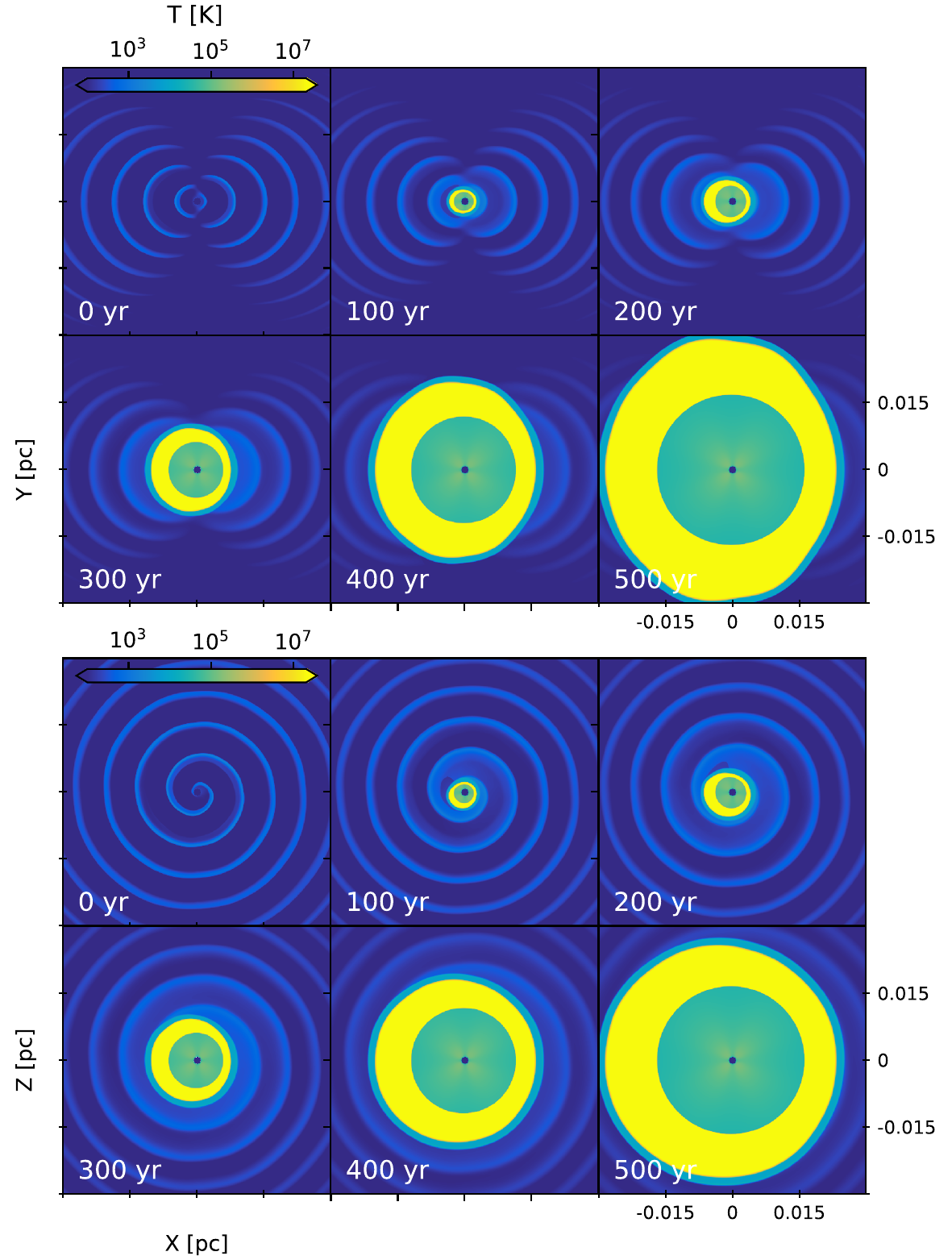}
    \caption{Temperature of the formation and evolution of a PN within a 3D spiral pattern of Model M2 (with binary period 558/2~yr) produced by a spherically-symmetric post-AGB wind (see Table~\ref{tab:models} for details). The top panels show the results for the $x-y$ plane whilst the bottom panels show those for the $x-z$ plane. The results are shown at times $t=0,100,200,300,400$ and $500$ yr after the onset of the post-AGB phase.}
\label{fig:pagb_period2_temp}
\end{figure}

In this appendix we present the temperature maps of the simulations presented in the main body of this article. Fig.~\ref{fig:pagb_period_temp} and  \ref{fig:pagb_period2_temp} show the creation of a hot bubble after the injection of an isotropic fast stellar wind. These figures are complementary to Fig.~\ref{fig:pagb_period} and \ref{fig:pagb_period2}.

\begin{figure*}
\centering
\includegraphics[width=0.8\linewidth]{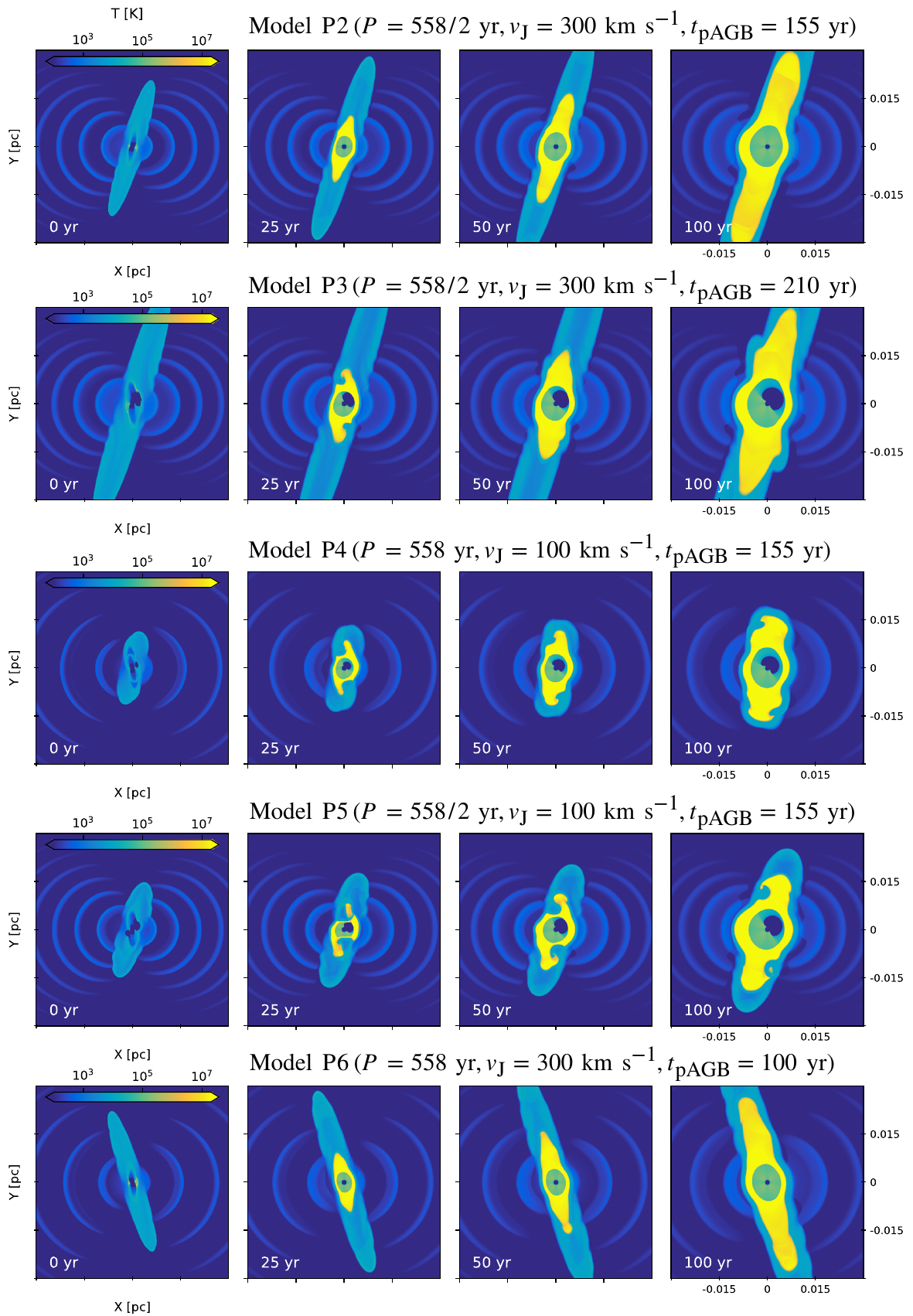} 
\caption{Evolution with time of the gas temperature $T$ for Models P2--P6 as seen in the $x-y$ plane. The results show the integration times $t=0,25,50$ and $100$ yr after the onset of the post-AGB phase. In all cases a jet has been injected and interacted with the 3D spiral pattern before the onset of the post-AGB phase. Details are listed in Table~\ref{tab:models} and described in Sec.~\ref{sec:post_AGB}.}
\label{fig:P2_temp}
\end{figure*}

Fig.~\ref{fig:P2_temp} shows the temperature evolution for simulations of bipolar PNe produced by jets. This figure represent the same snapshots presented in Fig.~\ref{fig:P2}.

\end{document}